\documentclass{aastex}
\usepackage{url}

\RequirePackage{color}

\begin{document}

\title{Are the magnetic fields of millisecond pulsars $\sim 10^{8}$ G?}
%\slugcomment{Not to appear in Nonlearned J., 45.}
%% Running heads
%\shorttitle{Short article title}
%\shortauthors{Autors et al.}

\author{Rafael S. de Souza} \and \author{Reuven Opher}
\affil{IAG, Universidade de S\~{a}o Paulo, Rua do Mat\~{a}o 1226, Cidade
Universit\'{a}ria, \\ CEP 05508-900, S\~{a}o Paulo, SP, Brazil}

%\email{\emaila}

%\altaffiltext{1}{IAG, Universidade de S\~{a}o Paulo, Rua do Mat\~{a}o 1226, Cidade
%Universit\'{a}ria, \\ CEP 05508-900, S\~{a}o Paulo, SP, Brazil.}

%\doublespacing

\begin{abstract}
It is generally assumed that the magnetic fields of millisecond pulsars (MSPs) are $\sim 10^{8}$G. We argue that this may not be true and the fields may be appreciably greater. We present six evidences for this: (1) The $\sim 10^{8}$ G field estimate is based on magnetic dipole emission losses which is shown to be questionable; (2) The MSPs in low mass X-ray binaries (LMXBs) are claimed to have $< 10^{11}$ G on the basis of a Rayleygh-Taylor instability accretion argument. We show that the accretion argument is questionable and the upper limit $10^{11}$ G may be much higher; (3) Low magnetic field neutron stars have difficulty being produced in LMXBs; (4) MSPs may still be accreting indicating a much higher magnetic field; (5) The data that  predict $\sim 10^{8}$ G for MSPs also predict ages on the order of,  and greater than,  ten billion years,  which is much greater than normal pulsars. If the predicted ages are wrong, most likely the predicted $\sim 10^{8}$ G fields of MSPs are wrong;   (6) When magnetic fields are measured directly with cyclotron lines in X-ray binaries, fields $\gg 10^{8}$ G are indicated.  Other scenarios should be investigated. One such scenario is the following. Over  85\% of MSPs are confirmed members of a binary. It is possible that all MSPs are in large separation binaries having  magnetic fields $> 10^{8}$ G  with their magnetic dipole emission being balanced by  low level  accretion from their companions. 
\end{abstract}


%\section*{}
%\label{sec:intro}

\section{Introduction}

The discovery of  millisecond pulsars (MSPs) was reported by \cite{bac82}. Their  origin was suggested to be due to spin-up by accretion torques of a binary companion \citep{alp82} such as in low mass X-ray binaries (LMXB).  The strength of the magnetic field of the MSPs is generally estimated assuming that they are    isolated  and there is no  accretion. Only  energy loss due to magnetic dipole  emission is  assumed. From the spin-down rate,  the magnetic field is estimated to be $\sim 10^{8}$ G, which is universally used.  All of the  articles published on the MSP since 1982 have assumed the validity of this  scenario and the  relatively low magnetic field $\sim 10^{8}$ G. This can be  compared with the typical $\sim 10^{12}$ G field of pulsars. In this article we argue that the magnetic fields of MSPs may be $> 10^{8}$ G. 

Assuming that the spin-down of a pulsar is due to magnetic dipole emission, the surface dipole magnetic field is obtained from the relation 
\begin{equation}
B = 3.2 \times 10^{19}(P\dot{P})^{1/2}~G, 
\end{equation}
where the neutron star moment of inertia is assumed to be $10^{45}~g ~cm^{2}$, the neutron star radius  $10^{16}~cm$, $P$ is the pulsar period in seconds, and $\dot{P}$ is its derivative \citep{man05}.

MSPs are characterized by spin-down rates four to six orders of magnitude less than normal pulsars implying ages $\sim 10^{10}$ years. It is argued that they probably require their short periods through a recycling process in which mass and angular momentum are transferred to an old and slowly rotating pulsar from a binary companion \citep{man04}.

We discuss in section II ``Can the magnetic field of MSPs be determined assuming magnetic dipole emission?", in section III ``Can the magnetic field of a MSP be determined by the rate of accretion in LMXBs?", in section IV ``Can low magnetic field MSPs be produced in LMXBs?", in section V ``Are MSPs still accreting?", in section VI  ``Are  the indicated ages of MSPs realistic?" and in section VII ``Measurement of magnetic fields in X-ray binaries with cyclotron lines". Our  ``Conclusions and Discussion" are presented in section VIII. 

\section{Can the magnetic field of MSPs be determined assuming magnetic dipole emission?}

In most theoretical models of pulsars the braking torque is proportional to some power, n, of the rotation frequency $\Omega ~(\equiv 2\pi/P)$
\begin{equation}
\dot\Omega = -K\Omega^{n}
\end{equation}
where $\dot\Omega$ is the time derivative of $\Omega$ \citep{man77}. The parameter n is known as the braking index and K is a positive constant. For braking by magnetic dipole radiation or particle acceleration in a dipole field, which is generally assumed,  n = 3  \citep{man77}. 

The very low spin-down rates of the MSPs have so far precluded any direct measurements of the braking index. An index of n = 3 is indicated for old radio pulsars, but values $n \sim 1.5-2.8$ have been measured in younger pulsars \citep{lyn96, hob04, fer07}.

 \cite{pac67} pointed out that a rotating magnetized star would radiate,  in the form of electromagnetic waves,  magnetic dipole radiation at the rotation frequency. The radiation reaction torque transmitted to the star by the magnetic field is 
\begin{equation}
N = -\frac{2(\mu \sin \alpha)^{2}}{3c^{3}}\Omega^{3}
\end{equation}
where $\mu$ is the magnetic-dipole moment and $\alpha$ is the angle between the magnetic and rotation axis.  \cite{gun69} and \cite{ost69}  recognized that this torque could account for the observed secular increase in pulsar periods. Since $N \propto \Omega^{3}$, the braking index n is three for these pulsars. Since $\mu \sim B_{0}R^{3}$, where $B_{0}$ is the surface magnetic field and $R$ is the radius of the star, the surface magnetic field  $B_{0}$ can be estimated from the observed period derivative (with $ \sin\alpha$ assumed  approximately equal to one)

\begin{equation}
B_{0} \approx \left(\frac{3Ic^{3}P\dot P}{8\pi^{3}R^{6}}\right)^{1/2}
\end{equation}
where $I$ is the moment of inertia. For $I \approx 10^{45}~gcm^{2}$, $R = 10^{6} ~cm$ and $P$ in seconds we obtain Eq. (1) (\citep{man77}).

It is not clear how an isolated MSP is formed, but
it is reasonable to assume that it is not found in a vacuum. A low
density plasma undoubtedly  exists around the MSP. In that case, the
electromagnetic dispersion relation has the form
\begin{equation}
\omega^{2} = \omega_{p,e}^{2}+c^{2}k^{2}
\end{equation}
 \citep[e.g.,][]{stu94} where
$\nu_{p,e}(=\omega_{p,e}/2\pi)$ is the plasma frequency, 
\begin{equation}
\nu_{p,e}=
10^{3.95}n_{e}^{1/2},
\end{equation}
and $n_{e}$ (in units of electrons per $cm^{3}$) is the electron density.  For $\omega < \omega_{pe}$, the wavenumber k is imaginary and the electromagnetic wave is rapidly damped. Thus an electromagnetic wave of frequency $\nu =
\omega/2\pi = 10^{3}$,  emitted  by a MSP,
will not propagate away from it  if the density of the
plasma is $n_{e} > 0.013$. This limit for the plasma
density is very low. It can be compared to n $\sim$ 1 for the
interstellar medium in our galaxy and $n\gg 1$ in a globular
cluster.

Particle acceleration by the rotating magnetic field of the neutron star was studied by \cite{gol69}. The rotation of the neutron star and its associated magnetic field produces strong electric fields in the space surrounding the star. \citeauthor{gol69} were the first to point out that,  because of these fields, the region surrounding the star cannot be a vacuum and must contain a substantial space charge \citep{man77}.

The work of \citeauthor{gol69} however, considered a simplified system (an axisymmetric rotator) whereas in fact pulsars must possess a non-axisymmetric field (an oblique rotator) in order to produce a periodic signal. Furthermore, they did not give  a self-consistent description of the currents and fields surrounding the star. Their model involves flow of charge of one sign through a region of space with charge of the opposite sign, a situation unlikely to exist in a real pulsar \citep{man77}.

From the above we can not be certain that the MSPs are radiating as a magnetic dipole and that the $P$ and $\dot P$  of MSPs give a true value of $B_{0}$.

\section{Can the magnetic field of a MSP be determined by the rate of accretion in LMXBs?}

It has also been argued that MSPs in LMXBs have low magnetic fields, $< 10^{11}$ G,  due to the accretion pressure produced by   a weak Rayleigh-Taylor instability. The argument is that,  if the field is higher, accretion onto the neutron star could not occur and produce the observed X-rays \citep{spr90,mil98,psa99,lam06}. If the stellar magnetic field is$~\gtrsim 10^{11}$G, all of the gas in the disk will  couple to the neutron star and be funneled out of the disk plane, toward the magnetic poles of the star. For  magnetic fields $< 10^{11}$G,  some of the accreting gas will continue as a disk flow towards the stellar surface as a result of the Rayleigh-Taylor instability and produce the observed X-rays. However, if the instability is greater than the Rayleigh-Taylor instability, magnetic fields $> 10^{11}$G  are possible. 

 Previous analysis neglected  the strongest disk instability, which is much greater than the Rayleigh-Taylor instability:  the magneto-rotational instability (MRI) \citep{bal91,bal98}. In a Keplerian disk the maximum growth rate of the MRI is $\sim (3/4) \Omega$,  with a typical perturbation wavenumber $\Omega/V_{A}$, where $\Omega (V_{A})$ is the angular (Alfven) velocity. The disk becomes very dynamic and full of reconnection events \citep{kud02,mac03}. 
 
 We thus conclude  that the  low accretion rate in LMXB could  occur for a stellar magnetic field$~\gtrsim  10^{11}$ G due to the MRI.

We discussed above one of the most popular scenarios for determining the magnetic field of a MSP by the rate of accretion  in LMXBs. Other means of accretion are possible. 
In the magnetic Cataclysmic Variables, where a magnetic white dwarf accretes mass from a low-mass companion star, the existence and extent of the disk strongly correlate with the strength of the magnetic field of the white dwarf. Hence, in the most strongly magnetic system ($10^7 - 10^8$ Gauss: the AM Herculis systems), accretion occurs directly via a funnel, without the formation of a disk. These strong fields are measured directly via the detection of strong cyclotron lines and/or Zeeman lines from the surface of the white dwarfs. In the Intermediate Polars, accretion occurs via a truncated disk, the extent of which is related to the strength of the field of the white dwarf. In some cases, fields in the Intermediate polars have also been measured, directly, via the detection of polarized radiation and, more recently, of
cyclotron lines in the polarization spectra.

\section{Can the low magnetic field MSPs be produced in LMXBs?}

It has been suggested that pulsar magnetic fields are lowered by recycling in binaries  (see \citeauthor{bha91} 1991 for a review). The mechanism for this remains unclear, with suggestions including decay of crustal fields due to heating \citep{blo86},  burial of the field \citep{bis74,rom90,rom93} and decay of core fields due to flux tube expulsion from the superfluid interior \citep{sri90}). \cite{kon97} suggested that rapid ohmic decay in the accretion heated crust occurs. On the one hand, the heating reduces the electrical conductivity and consequently the ohmic decay time-scale induces a faster decay of the field. On the other hand, the deposition of matter on top of the crust pushes the original current carrying layers into deeper and denser regions where the higher conductivity slows down the decay \citep{kon97}.

In a class of recycling models in which the magnetic field decrease is a function only of the amount accreted onto the neutron star,  it was shown that no model of this class is consistent with all available data \citep{wij97}.

The detection of coherent X-ray pulsations with a millisecond period in a handful of LMXBs \citep{lam05} is often used in support of the idea of accretion-induced field decay \citep{wij98}. However, whether this is evidence simply for field submersion and spin-up during the accretion disk phase, or for field decay and spin-up, remains to be established \citep{fer07}. 

We conclude from the above, that it is not clear if  low field $\sim 10^{8}$ G field MSPs can be produced in LMXBs. 

We pointed out that the field could simply be submerged by the accreted matter and would then re-emerge later on when accretion stops. \cite{zha09} show that there is no evidence for field restructuring and/or decay in accreting magnetic white dwarfs. This fact was already known for the isolated magnetic white dwarfs. As far as we know, this could apply also to neutron stars.

\section{Are MSPs still accreting?}

 Almost all $(>85\%)$ MSPs have been confirmed to be in binaries. The central idea behind the MSP-LMXB connection is that LMXBs can provide the long-lived phase of moderate mass transfer rates thought necessary to spin-up the neutron star to millisecond periods. The MSP is supposed to be produced when the accretion shuts off.

But does the accretion ever shut-off? In order to have a shut-off of the mass accretion rate,  $\dot M$, the accretion evolution time 
\begin{equation}
\tau_{\dot M} \equiv \frac{\dot M}{\dot M^{'}}
\end{equation}
needs to be shorter than the spin-down time scale 
\begin{equation}
\tau_{spin} \equiv \frac{P_{spin}}{\dot P_{spin}} \propto \dot M^{-3/7}
\end{equation}
or
\begin{equation}
\tau_{\dot M} < \tau_{spin}.
\end{equation}

The accretion rate, $\dot M$, decreases as the orbital separation grows. At late times, for gravitational wave driven orbital evolution in the limit when the low mass companion of the neutron star is very much less than that of the neutron star, 

\begin{equation}
\tau_{\dot M} \propto \dot M^{-11/14}
\end{equation}
Thus as $\dot M$ decreases, the ratio $\tau_{spin}/\tau_{\dot M}$ decreases. For a neutron star with a magnetic field $B = 5 \times 10^{8}$ G, the two timescales are roughly equal at $\dot M \sim 10^{-9} M_{\odot}/yr$. The system thus has time to come into equilibrium and accretion never shuts off \citep{del08}.

 It was noted by \cite{del08} that
in two of the four classes of LMXBs, mass transfer to the neutron star never shuts off. Both
of these systems have gradually declining mass transfer rates, as the
system evolves to longer orbital periods. \citeauthor{del08} noted that all known MSPs can be associated with one
of these two classes. 

Thus,  MSPs may  always be subject to
a very low accretion rate.

\section{Are the indicated ages of MSPs realistic?}

In normal pulsars $P$ and $\dot P$ are distributed about mean values $P\sim 0.1 s$  and $\dot P \sim 10^{-5} s s^{-1}$. This implies fields $B \sim 10^{11}-10^{13}$ G. Values of MSPs of $P \sim 5 ms$ and $\dot P \sim 10^{-20} s s^{-1}$ imply fields of $ B \sim 10^{8}-10^{9}$ G. The age of the MSP is obtained from the relation 
\begin{equation}
\tau \equiv \frac{P}{(n-1)\dot P}\left[1-\left(\frac{P_{0}}{P}\right)^{n-1}\right]
\end{equation}
where  $P_{0}$ is the initial period. For $P_{0} \ll P$ and n = 3
\begin{equation}
\tau = \frac{P}{2\dot P}
\end{equation}

From this expression, however, the ages of many MSPs are more than ten billion years \citep{fer07}.
Half of the corrected sample of MSPs of \cite{tos99} have characteristic ages comparable or greater than ten billion years \citep{fer07}.

These ages are orders of magnitude greater than the ages of normal pulsars. Since the ages of the MSPs  are obtained from $P$ and $\dot P$ assuming the scenario that magnetic dipole emission is the only energy loss mechanism, and there is no accretion,  a doubtful age makes the scenario doubtful and the evaluation of $B_{0}$ from $P$ and $\dot P$ doubtful.

\section{Measurement of magnetic fields in X-ray binaries with cyclotron lines}

 The only sure way of measuring magnetic fields is by observing the electron-cyclotron line. The many observed electron-cyclotron measurements of magnetic fields in X-ray binaries indicate fields $\gg 10^{8}$G: \textbf{4U 0352+309} \cite{cob02,del01}, \textbf{GX 301-2} \cite{cob02}, BATSE Pulsar team\footnote{http://www.batse.msfc.nasa.gov/batse/pulsar/index.html }, \textbf{4U 1538-52} \cite{cob02}, BATSE Pulsar team, \textbf{4U 1907+09} \cite{cob02,bay01}, \textbf{Vela X-1}\cite{cob02}, BATSE Pulsar team; \textbf{A 0535+26} \cite{neg00}, \textbf{XTE J19646+274}
\cite{cob02}, \textbf{LMC X-4} \cite{lab01}, \textbf{4U 1626-67} \cite{cob02,van98}, BATSE Pulsar team, \textbf{Cen X-3} \cite{cob02,nel97},  BATSE Pulsar team; \textbf{V 0332+53} \cite{shi02}; \textbf{4U 0115+63} \cite{cob02}, \textbf{Her X-1} \cite{cob02,oos01},  BATSE Pulsar team. 
No cyclotron measurement has indicated a magnetic field $\sim 10^{8}$ G on the neutron star. 

To be complete, the lack of detection of cyclotron lines in the X-ray spectra of LMXBs could be used, however, to support the view that their fields are low. 
The fields measured in accreting neutron stars using cyclotron lines are those of neutron stars in High Mass X-Ray Binaries, which are still very young and mostly wind accretors. If the fields of neutron stars in LMXRBs were of the order of $10^8$ Gauss, the cyclotron lines would be in the optical spectral band. Such lines, arising from extremely small accretion shocks, would be swamped by the radiation from the accretion disk and would thus be hidden from detection.

\section{Conclusions and Discussion}

The period $P$ and its time derivative $\dot P$ are used to obtain the magnetic field $B_{0} \sim 10^{8}$ G of MSPs.  We presented six  evidences in sections II-VII which indicate that the measured $P$ and $\dot P$ may not indicate the true value of $B_{0}$ and that $B_{0}$ may be $> 10^{8}$ G. The evidences are:

\begin{enumerate}
\item Magnetic dipole emission is assumed with a braking index n = 3. The braking index of MSPs, however, have never been measured. We show that the magnetic dipole electromagnetic emission does not propagate in realistic interstellar media and a self-consistent model of magnetic dipole particle emission has never been made, putting doubt on the necessity of the  magnetic dipole emission assumption for MSPs;
\item It is assumed that in LMXBs only the Rayleigh-Taylor instability occurs in the accretion disks and a magnetic field $< 10^{11}$ G needs to be present at the surface of the neutron star to allow for accretion to occur. The much stronger magneto-rotational instability is probably present allowing for a field $\gg 10^{11}$ G on the neutron stars;

\item It has been assumed that the abnormal low magnetic fields $\sim 10^{8}$ G for pulsars could be produced  in LMXBs. There is, however, no self-consistent theory that does this;

\item The standard  scenario is that the pulsar is spun-up by accretion and then the accretion stops leaving the MSP to spin-down  only by magnetic dipole emission.  The accretion may never stop, however. A balance between  dipole emission  balanced by accretion could occur;

\item The $P$ and $\dot P$ of MSPs indicate ages $\gtrsim 10^{10}$ years. This is comparable to  the age of the Universe and very much older than other pulsars; and 

\item Measurements of cyclotron lines in X-ray binaries all indicate magnetic fields $\gg 10^{8}$G.

\end{enumerate}

Together, the six above evidences indicate that the general scenario that is used to deduce from $P$ and $\dot P$ of MSPs that $B_{0} \sim 10^{8}$ G may be incorrect. We suggest  investigating alternate scenarios. A possibility is that  MSPs may have $B \gg 10^{8}$ G and that the observed $\dot P$ of MSPs is due to a balance of magnetic dipole emission with a low-level of  accretion from a companion in a large separation binary. 

To be complete, the magnetic fields of MSPs, however, may be   low due to the following scenario.  The neutron stars were weakly magnetic at birth (maybe as a consequence of AIC). In this case, field decay would not be necessary. After all, only about 15\% of white dwarfs are magnetic, with magnetic fluxes similar to those inferred for radio pulsars. It is not clear why one should assume that ALL neutron stars are highly magnetic.

\acknowledgments
R.S.S. thanks the Brazilian agency FAPESP for financial support
(2009/06770-2). R.O. thanks FAPESP (06/56213-9) and the Brazilian
agency CNPq (300414/82-0) for partial support. We would like to thank an unknown referee for very helpful comments. His comments were included at the end of sections 3, 4, 6 and 8.

%\bibliographystyle{spr-mp-nameyear-cnd}
%\bibliography{ref}

\end{document}